\documentclass[
]{ceurart}

\usepackage{listings}
\usepackage{algorithm}
\usepackage{algpseudocode}
\usepackage{makecell}
\usepackage{multirow}
\usepackage{subcaption}
\usepackage{url}
\usepackage{helvet}
\usepackage{courier}
\usepackage{tabularx}  
\usepackage{array}
\usepackage{booktabs}
\usepackage{tcolorbox}

\begin{document}

\copyrightyear{2025}
\copyrightclause{Copyright for this paper by its authors.
  Use permitted under Creative Commons License Attribution 4.0
  International (CC BY 4.0).}

\conference{AIIDE Workshop on Experimental AI in Games(EXAG),
  Nov 10--14, 2025, University of Alberta, Edmonton, AB, Canada}

\title{Narrative-to-Scene Generation: An LLM-Driven Pipeline for 2D Game Environments}


\author[1]{Yi-Chun Chen}[%
orcid=0009-0003-4035-9894,
email=ychen74@alumni.ncsu.edu,
url=https://sites.google.com/view/rimichen-web/bio,
]
\fnmark[1]
\address[1]{Yale University, New Haven, CT 06510, US}

\author[2]{Arnav Jhala}[%
orcid=0000-0003-3874-1720,
email=ahjhala@ncsu.edu,
url=https://www.csc.ncsu.edu/people/AHJHALA,
]
\address[2]{North Carolina State University, Raleigh, NC 27606, US}

\cortext[1]{Corresponding authors.}

\begin{abstract}
Recent advances in large language models (LLMs) enable compelling story generation, but connecting narrative text to playable visual environments remains an open challenge in procedural content generation (PCG). We present a lightweight pipeline that transforms short narrative prompts into a sequence of 2D tile-based game scenes, reflecting the temporal structure of stories. Given an LLM-generated narrative, our system identifies three key time frames, extracts spatial predicates in the form of "Object-Relation-Object" triples, and retrieves visual assets using affordance-aware semantic embeddings from the GameTileNet dataset~\cite{chen2025gametilenet}. A layered terrain is generated using Cellular Automata, and objects are placed using spatial rules grounded in the predicate structure. We evaluated our system in ten diverse stories, analyzing tile–object matching, affordance–layer alignment, and spatial constraint satisfaction across frames. This prototype offers a scalable approach to narrative-driven scene generation and lays the foundation for future work on multi-frame continuity, symbolic tracking, and multi-agent coordination in story-centered PCG.
\end{abstract}


\maketitle

\section{Introduction}

Large language models (LLMs) can generate rich, coherent narratives from minimal input, creating new opportunities for procedural content generation (PCG) in games that emphasize narrative and visual storytelling. Yet while narrative generation has advanced rapidly, translating stories into structured, spatially grounded game scenes remains an open challenge. Most PCG systems still produce isolated levels or static layouts without modeling temporal structure and continuity—core aspects of narrative experiences that connect space, time, and character action.

Narratives unfold through sequences of events that shape spatial progression and thematic coherence~\cite{mccloud1993understanding,cohn2013visual}. Building on this idea, we present a method for generating sequences of visual game scenes from short narratives. Each story is segmented into three temporal frames (beginning, middle, end) following narrative-structure conventions such as Freytag’s pyramid, capturing key transitions while keeping generation tractable. This segmentation supports partial modeling of story progression and provides a foundation for studying temporal consistency and dynamic storytelling.

Our approach introduces a symbolic-to-visual pipeline that extracts spatial predicates from each frame, maps them to assets in the GameTileNet dataset~\cite{chen2025gametilenet}, and arranges them on procedurally generated terrains. Object–relation–object triples are derived through LLM prompting and grounded in visual tiles using semantic and affordance-based filtering. Terrain layers are produced with Cellular Automata to ensure connectivity, while objects are placed through rule-based spatial refinement that satisfies adjacency and containment constraints. The resulting layered 2D scenes reflect the structure of the source narrative while maintaining spatial logic relevant to gameplay.

Our contribution lies not in proposing new generative algorithms but in integrating established PCG components, such as Cellular Automata and semantic matching, into a unified narrative-to-scene pipeline emphasizing temporal segmentation and affordance grounding. This coordination supports cross-frame continuity and balances semantic fidelity with affordance validity, a combination rarely explored in prior PCG work. LLM-generated stories serve only as a reproducible testbed for open-vocabulary narrative input; the pipeline itself is agnostic to text source and can process any authored story.

\begin{table*}[!ht]
\caption{Example of generating a tile-based game scene from narrative descriptions. The system extracts spatial relations and semantic objects (left), selects corresponding tile images from the GameTileNet dataset (middle), and renders a visual scene that satisfies both semantic and spatial constraints (right). Top: labeled object layout. Bottom: rendered tile-mapped scene.}
\label{teaser}
\centering
\renewcommand{\arraystretch}{1.1}
\setlength{\tabcolsep}{3pt}
\begin{tabular}{|
  m{3.5cm}|
  m{4.2cm}|
  >{\centering\arraybackslash}m{5.2cm}|}
\hline
\textbf{Scene Description} & \textbf{Matched Tiles} & \textbf{Rendered Scene} \\
\hline
\begin{minipage}[c]{\linewidth}
\centering
\textcolor{red}{House} below \\
\textcolor{green}{Tree} \\
\textcolor{green}{Tree} to the right of \textcolor{brown}{Barrel} \\
\textcolor{yellow}{Flower} above \\
\textcolor{green}{Tree} \\
\textcolor{orange}{Tree stump} to the left of \textcolor{green}{Tree}
\end{minipage}
&
\begin{minipage}{\linewidth}
\centering
\setlength{\tabcolsep}{2pt}
\begin{tabular}{cc}
\includegraphics[height=1cm]{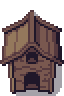} & 
\includegraphics[height=1cm]{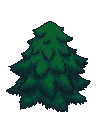} \\
House & Tree \\
\end{tabular}\\[1pt]
\begin{tabular}{cc}
\includegraphics[height=1cm]{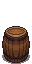} & 
\includegraphics[height=1cm]{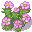} \\
Barrel & Flower \\
\end{tabular}\\[1pt]
\begin{tabular}{c}
\includegraphics[height=1cm]{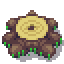} \\
Tree stump \\
\end{tabular}
\end{minipage}
&
\begin{minipage}[c]{0.95\linewidth}
\centering
\vspace{4pt}
\includegraphics[height=0.8\linewidth]{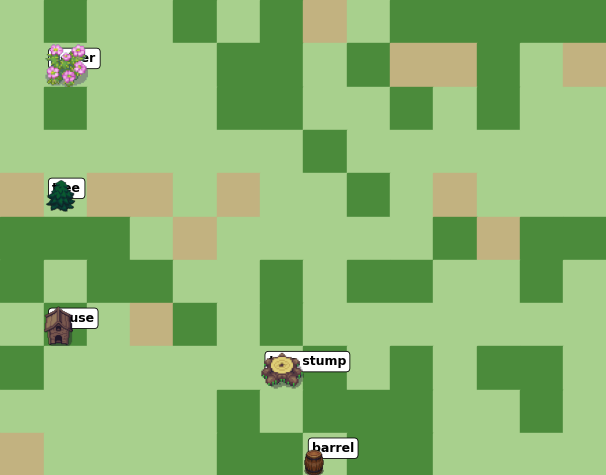}
\vspace{4pt}
\end{minipage}
\\
\hline
\end{tabular}
\end{table*}

We evaluate the system on ten narrative examples, each divided into three frames. Evaluation considers semantic-tile matching, affordance alignment, and spatial-relation satisfaction in the generated scenes. Even with lightweight symbolic rules and open-ended input, the results exhibit coherent spatial organization that reflects narrative intent.

\textbf{Contributions.}
\begin{itemize}
    \item A unified pipeline that decomposes short narratives into temporally segmented, layered 2D scenes.
    \item A predicate-extraction and semantic-matching process that grounds narrative objects in tile-based assets using affordance-aware filtering.
    \item A multi-layer scene synthesis method combining Cellular Automata terrain generation with rule-based spatial placement.
    \item An empirical analysis of ten narrative examples demonstrating semantic alignment, affordance coherence, and spatial-relation satisfaction across frames.
\end{itemize}

The proposed system offers a modular foundation for narrative-to-scene generation, emphasizing temporal segmentation, affordance grounding, and semantic alignment. Although the current prototype does not model agent behavior or dynamic state transitions, it establishes a framework extendable to multi-agent coordination and gameplay-aware narrative PCG. Recent work shows that LLMs can segment narrative events in ways that align with human perception~\cite{michelmann2025large}, supporting our choice to use structured time frames as the basis for visual scene construction. Table~\ref{teaser} illustrates one such example, where spatial constraints and tile affordances jointly produce semantically coherent game scenes from narrative descriptions.

\section{Related Work}

\subsection{Narrative Theory and Structure in Games}
Narrative structure in games has been studied through formalist, experiential, and emergent perspectives. Early work examined the tension between linear storytelling and player agency \cite{aarseth2012narrative, juul2005games}. Later models addressed experiential modes \cite{calleja2009experiential} and story architectures that accommodate agency \cite{domsch2013storyplaying}. Other work connected narrative to spatial layout and symbolic environments \cite{domsch2019space, lindley2005story}, while surveys summarize recurring patterns in interactive media \cite{koenitz2024narrative}. These theories provide foundations for incorporating narrative intent into generative pipelines.


\subsection{Narrative-Guided Procedural Content Generation}

Procedural content generation (PCG) increasingly leverages narrative structure. Quest and story grammars guide goal-oriented generation~\cite{howard2022quests}, and language models now produce worlds and events~\cite{nasir2024word2world,todd2024gavel}. Emotional arcs and planning constraints improve coherence and engagement~\cite{miller2019stories}. PCG methods span grammars, search-based strategies, and PCGML~\cite{summerville2018procedural,yannakakis2025procedural}, with reinforcement learning extending control to narrative form~\cite{togelius2010search}. Layered generation has also appeared in visual storytelling and comics~\cite{chen2024collaborative,chen2023customizable}, demonstrating ways to link symbolic story input with generative models. Prior text-to-scene and PCG systems generate single scenes or levels~\cite{riedl2010narrative,guzdial2018combinatorial,summerville2018procedural} but rarely model temporal continuity or explicit narrative structure. Our work addresses this gap by connecting narrative segmentation with spatial and affordance reasoning, bridging story progression and playable spatial layouts.


\subsection{Tile-Based Generation and Semantic Affordances}
Tile-based abstractions remain central to scalable generation. Representative methods include constrained layout via SMT solvers \cite{whitehead2020spatial}, evolutionary search \cite{petrovas2022procedural}, and grammar-based dungeon layouts \cite{jiang2021synthesizing}. Neural approaches introduce embeddings for generalized level generation \cite{jadhav2021tile} and segmentation for context-aware tilemaps \cite{gabriel2023semantic}. Affordance modeling distinguishes environmental, interactive, and collectible roles \cite{hunicke2004mda}, while corpora enable affordance-aware generation \cite{bentley2019videogame}. GameTileNet contributes a dataset of low-resolution tiles annotated with layered affordances \cite{chen2025gametilenet}.

\subsection{Semantic Matching and LLM-Guided Grounding}
LLMs are increasingly used for scene synthesis and grounding. Prompting and multimodal alignment allow them to act as planners for layouts \cite{volum2022craft, zhou2024scenex}. Studies show that structured input such as scene semantics improves alignment \cite{cao2023comprehensive}. Visual-semantic reasoning frameworks for matching and embedding inform techniques for aligning text predicates with tiles \cite{li2019visual, li2022image}. Cognitive studies suggest LLMs approximate human-like segmentation of narrative events, underscoring their potential for symbolic-neural integration \cite{michelmann2025large}.

\subsection{Symbolic Structures in Visual Storytelling}
Symbolic and graph-based representations provide structure for storytelling. Knowledge-enhanced generation has been modeled with narrative graphs or multimodal scene graphs \cite{hsu2020knowledge, xu2021imagine, mishra2019storytelling}. Other frameworks explore relational encodings to ensure continuity and causality in stories and games \cite{rivera2024story, akimoto2017computational}. Recent work highlights hierarchical narrative graphs as interpretable intermediaries for multimodal alignment \cite{chen2025hierarchical}. These approaches underscore the importance of symbolic scaffolds for controllable and interpretable storytelling pipelines.

\section{Method: Narrative-to-Scene Generation Pipeline}

\begin{figure*}[htbp]
    \centering
    \includegraphics[width=0.9\textwidth]{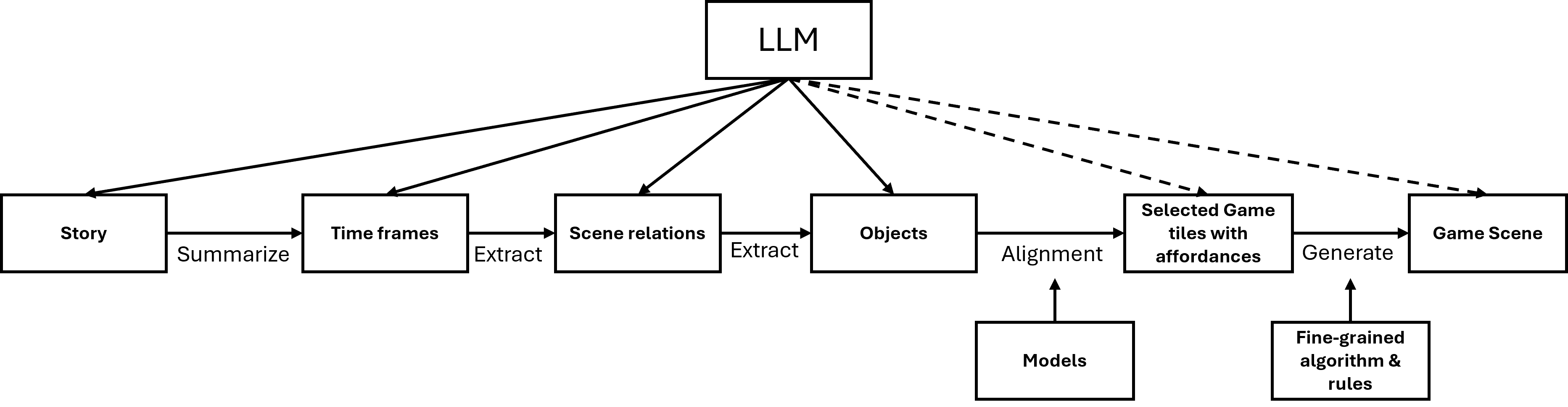}
    \caption{Pipeline for generating a tile-based game scene from an LLM-generated narrative. 
    Modules include narrative parsing, semantic matching, procedural terrain generation, and layered scene rendering.}
    \label{fig:processing_pipeline}
\end{figure*}

We propose a structured pipeline that transforms narrative text into visually grounded, tile-based game scenes. 
As illustrated in Figure~\ref{fig:processing_pipeline}, the pipeline integrates large language model (LLM)–based story generation with a sequence of symbolic and visual reasoning modules. 
The process begins with narrative prompting and temporal abstraction, where the story is segmented into key time frames and represented as predicate-style triples. 
These structured descriptions serve as input for semantic reasoning and symbolic grounding. 
Although the LLM provides the narrative input, this step serves only to ensure reproducible open-vocabulary test cases; the pipeline itself is model-agnostic and can process any authored story with minimal preprocessing.

Each scene is synthesized through a layered generation process: (1) terrain generation using Cellular Automata ensures navigable base regions; (2) semantic object matching retrieves tile assets aligned with narrative entities based on name, category, and affordance embeddings; and (3) spatial placement with rule-based refinement positions entities according to annotated spatial relations. 
The final output consists of rendered tile-based scenes and underlying semantic maps, enabling both human-readable visualization and downstream interactive applications.

\subsection{Narrative Prompting and Frame Extraction}

Inspired by \textit{Word2World}~\cite{nasir2024word2world}, we use a large language model to generate a short adventure story and extract three key narrative moments. 
Each moment is structured as symbolic predicates describing relationships between story elements, represented as \texttt{[Object] [Relation] [Object]} triples. 
These structured predicates form the basis for downstream spatial reasoning and visual scene generation.

The process uses two prompts:
\begin{itemize}
    \item \textbf{Prompt\_1:} ``Generate a short adventure story (about 100 words).''
    \item \textbf{Prompt\_2:} ``Extract three key time frames and describe each with [Object] [Relation] [Object] triplets.''
\end{itemize}

\vspace{4pt}
\noindent\textbf{Generated Story Example:}
\begin{tcolorbox}[colback=gray!5!white, colframe=gray!60!black, boxrule=0.5pt, arc=3pt]
\small
\textit{
In the heart of the Enchanted Forest, young Elara discovered an ancient map hidden within a hollow oak. 
It led her to the legendary Crystal Cavern, rumored to grant the finder a single wish. 
Braving treacherous paths and wild creatures, Elara reached the cavern's shimmering entrance. 
Inside, she faced the Guardian, a majestic dragon. 
With courage and wit, she solved the Guardian’s riddle, earning her the wish. 
Elara wished for peace in her war-torn village. 
As she exited the cavern, the skies cleared, and harmony was restored, proving that bravery and hope could transform the world.
}
\end{tcolorbox}
\vspace{6pt}

Table~\ref{table:time_frames} shows the extracted time frames, the corresponding predicate triples, and the tile-based scenes rendered by our pipeline.

\begin{table*}[ht!]
\caption{LLM-generated story, predicate triples, and corresponding rendered scenes for three narrative time frames.}
\label{table:time_frames}
\centering
\begin{tabular}{|c|c|c|}
\hline
\textbf{Time Frame 1} & \textbf{Time Frame 2} & \textbf{Time Frame 3} \\ \hline
Elara discovers the ancient map & Elara faces the treacherous paths & Elara meets the Guardian dragon \\ 
\hline
\makecell{
    \textcolor{green}{Hollow oak} contains \\
    \textcolor{red}{ancient map} \\
    \textcolor{brown}{Elara} stands near\\
    \textcolor{green}{hollow oak}\\
    \textcolor{yellow}{Sunlight} filters through\\
    \textcolor{green}{forest canopy}
    }
&
\makecell{
    \textcolor{brown}{Elara} walks along \\
    \textcolor{gray}{rocky path} \\
    \textcolor{orange}{Wild creatures} hide behind\\
    \textcolor{green}{dense bushes}\\
    \textcolor{yellow}{Treacherous paths} lead to\\
    \textcolor{gray}{Crystal Cavern}
    }
& 
\makecell{
    \textcolor{gray}{Crystal Cavern entrance} glows with \\
    \textcolor{yellow}{shimmering light} \\
    \textcolor{orange}{Guardian dragon} sits atop\\
    \textcolor{magenta}{crystal throne}\\
    \textcolor{brown}{Elara} stands before\\
    \textcolor{orange}{Guardian dragon}
    }
\\
\hline
\includegraphics[width=0.3\textwidth]{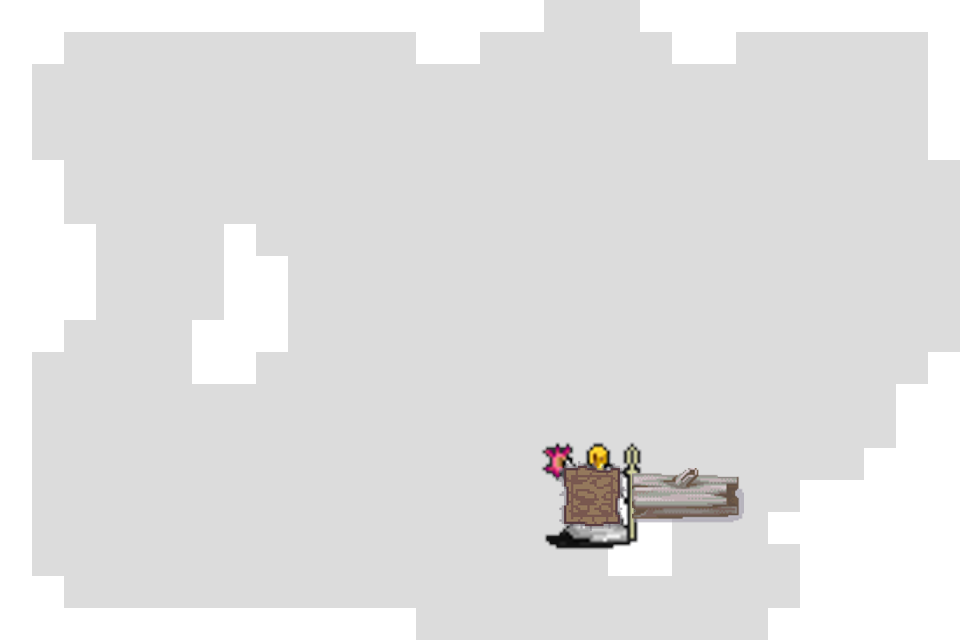}
&
\includegraphics[width=0.3\textwidth]{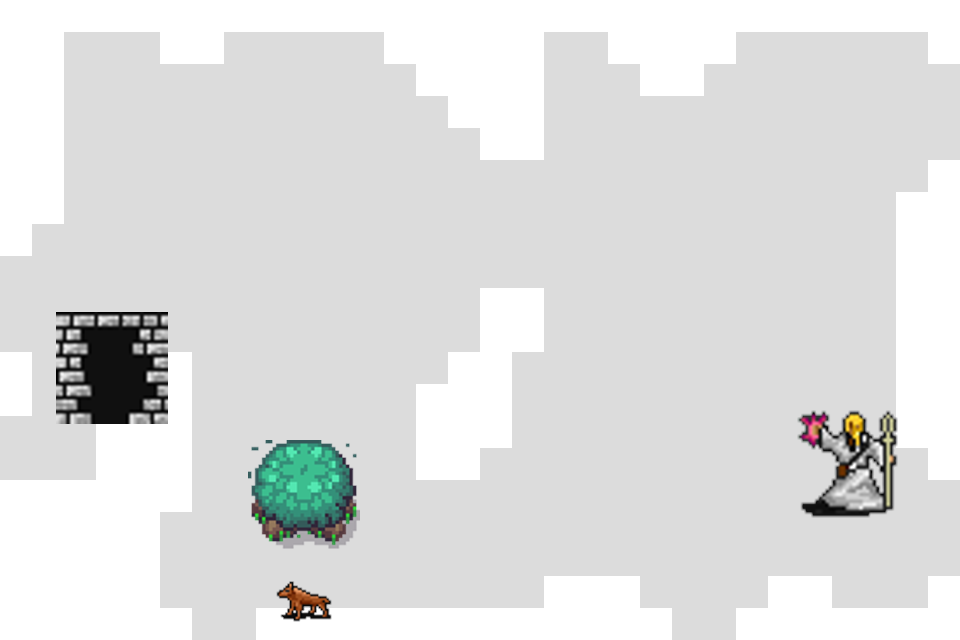}
&
\includegraphics[width=0.3\textwidth]{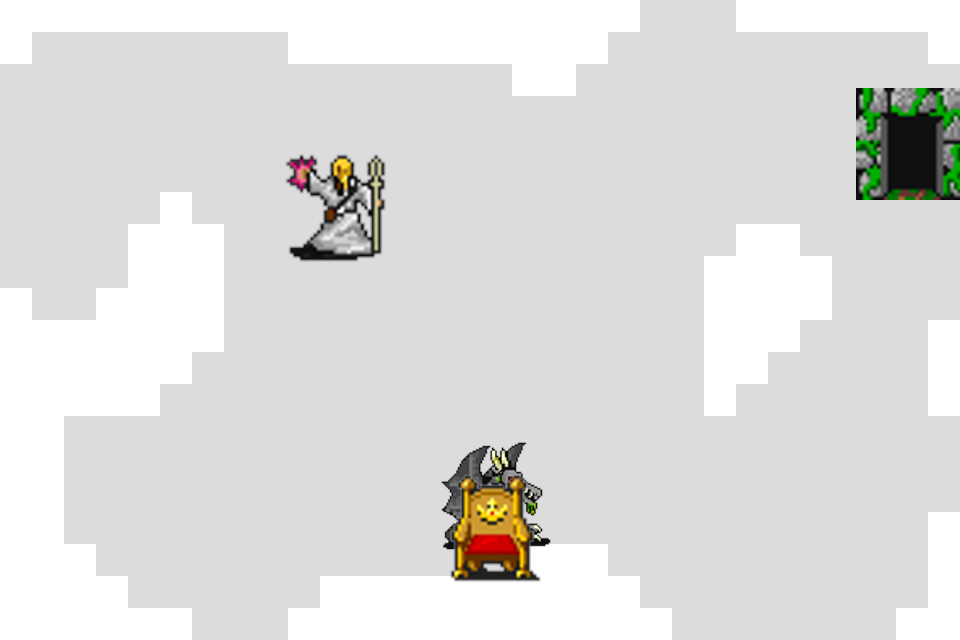}
\\ 
\hline
\end{tabular}
\end{table*}

\subsection{Symbolic Spatial Relation Mapping}

To transform symbolic narrative structure into spatial configurations, we begin by parsing each scene into a set of predicate triples in the form \texttt{[Object] [Relation] [Object]}. These relations encode spatial intent, such as adjacency or containment, and are mapped to a controlled ontology of canonical spatial actions suitable for tile-based rendering.

This mapping is informed by prior work on relational scene representation~\cite{krishna2017visual,johnson2017clevr}, which demonstrates that spatial relations such as \textit{above}, \textit{next to}, and \textit{on top of} align with how people describe and interpret visual layouts. For our system, we adopt the following spatial relation types:

\begin{itemize}
    \item \textbf{above / below:} vertical adjacency (Y-axis offset)
    \item \textbf{at left of / at right of:} horizontal adjacency (X-axis offset)
    \item \textbf{on top of:} overlapping placement with layer prioritization
\end{itemize}

Because natural language varies widely, we use a large language model (LLM) to normalize open-ended expressions to this spatial ontology. For example, \textit{contains} is mapped to \textit{on top of}, while \textit{stands near} may correspond to \textit{at left of} or \textit{at right of} depending on context. Human verification ensures the consistency and interpretability of the mappings. The resulting spatial relations serve as symbolic scaffolds that guide object placement during scene generation.

\subsection{Semantic Asset Retrieval with GameTileNet}

To align narrative objects with appropriate visual tiles, we adopt a semantic embedding–based retrieval strategy grounded in the GameTileNet dataset~\cite{chen2025gametilenet}. Each tile in the dataset is annotated with structured metadata, including object name, group label, supercategory, and affordance type. These attributes are embedded using the \texttt{all-MiniLM-L6-v2} Sentence Transformer to construct a searchable index. Narrative objects are encoded using the same model, and tile matches are retrieved via cosine similarity.

\paragraph{Affordance Types.}
GameTileNet classifies tiles into five affordance types adapted from the Video Game Description Language (VGDL)~\cite{schaul2013video}:

\begin{itemize}
  \item \textbf{Terrain:} Walkable ground surfaces (e.g., grass, stone).
  \item \textbf{Environmental Object:} Static scene elements (e.g., trees, fences).
  \item \textbf{Interactive Object:} Triggerable or functional elements (e.g., doors, levers).
  \item \textbf{Item/Collectible:} Usable or acquirable items (e.g., potions, scrolls).
  \item \textbf{Character/Creature:} Playable or non-playable agents (e.g., goblins, shopkeepers).
\end{itemize}

Affordance labels serve as soft constraints to improve retrieval robustness, especially for ambiguous cases. For example, the term "guardian" could refer to a statue or a creature, and the affordance context helps disambiguate the intended match. This retrieval process enables semantic alignment between narrative elements and visual assets while respecting scene composition constraints.

\subsection{Terrain and Scene Layout}

To render each scene with appropriate environmental context, we infer base terrain types and subregion patches from narrative content. Each scene is structured as a layered grid, with the base layer representing walkable terrain and additional layers corresponding to object affordance types.

\subsubsection{Terrain Suggestion via LLM Classification}

We use an LLM-based classification step to extract environmental cues from narrative objects. Following story decomposition into predicate triples, each object is assigned:

\begin{itemize}
    \item \textbf{Affordance type:} One of terrain, environmental object, interactive object, item/collectible, or character/creature.
    \item \textbf{Suggested terrain:} A free-text label describing the implied environment (e.g., "forest", "desert").
\end{itemize}

These predictions are aggregated to determine dominant terrain types for each scene.

\subsubsection{Base and Patch Selection with Continuity Propagation}

To ensure continuity across time frames, we assume that scenes with no explicit location change remain in the same environment. Scenes are grouped based on inferred location continuity using temporal adjacency and terrain similarity. For each group:

\begin{itemize}
    \item The most frequent terrain type is assigned as the \textbf{base terrain}.
    \item Objects with terrain-related labels (e.g., "path", "alley") are extracted as \textbf{patch candidates}.
    \item Patch terrain decisions are propagated across all scenes within the group.
\end{itemize}

This process maintains visual and narrative coherence across sequential scenes. The terrain selection process is summarized in Algorithm~\ref{alg:terrain_inference}.

\begin{algorithm}[H]
\caption{InferBaseAndPatchTerrain}
\label{alg:terrain_inference}
\begin{algorithmic}[1]
\Require story\_scenes: list of scenes with narrative objects
\Ensure base\_terrains: map from scene to base terrain label \\
\hspace{4.5em} patch\_terrains: map from scene to list of patch labels

\State Initialize empty maps: base\_terrains, patch\_terrains
\State Group scenes by inferred location continuity
\ForAll{scene\_group in grouped scenes}
    \State Initialize frequency\_counter
    \ForAll{scene in scene\_group}
        \ForAll{object in scene.objects}
            \State Predict affordance and suggested terrain using LLM
            \If{object.affordance == Terrain}
                \State Increment frequency\_counter[object.suggested\_terrain]
            \ElsIf{object.name contains terrain keywords}
                \State Append object.suggested\_terrain to patch\_terrains[scene]
            \EndIf
        \EndFor
    \EndFor
    \State base\_terrain $\gets$ terrain with max frequency in frequency\_counter
    \ForAll{scene in scene\_group}
        \State base\_terrains[scene] $\gets$ base\_terrain
    \EndFor
\EndFor
\end{algorithmic}
\end{algorithm}

\subsubsection{Scene Initialization with Cellular Automata}

We generate the layout of the base terrain using a Cellular Automata (CA)–based synthesis process. For each scene:

\begin{itemize}
    \item A connected walkable region is created using CA and verified for reachability.
    \item Terrain patches are inserted as subregions constrained within the generated base mask.
    \item The final output is a layered map suitable for narrative-aligned object placement.
\end{itemize}

This multi-layered scene layout ensures compatibility between narrative framing and spatial structure.







\subsection{Spatial Constraint--Driven Object Placement}

After terrain generation and semantic matching, each scene is populated by placing narrative-aligned objects within a multi-layer tile grid. This stage is divided into two parts: random initialization and symbolic refinement.

\subsubsection{Initial Placement on Walkable Terrain}
Objects are first placed randomly on the walkable base terrain using a greedy assignment process. For each object:
\begin{itemize}
    \item A walkable coordinate is selected from the base terrain mask.
    \item The object is placed into the corresponding layer based on its affordance (character, item, interactive, or environment).
\end{itemize}

\subsubsection{Spatial Relation–Based Refinement}
Once initial placements are made, we apply symbolic spatial constraints derived from narrative predicates. For each predicate of the form \texttt{[Object A] [Relation] [Object B]}, we apply the associated spatial transformation to reposition Object A relative to Object B. We define a rule-based adjustment engine to implement these spatial relations. Each relation is translated into a spatial offset and applied iteratively.

\begin{algorithm}[H]
\caption{ApplySpatialRelations}
\label{alg:apply_spatial}
\begin{algorithmic}[1]
\Require scene: object placement layers and predicate relations
\Ensure updated object positions satisfying spatial constraints

\ForAll{relation in scene.spatial\_relations}
    \State $(A, R, B) \gets$ relation.source, relation.relation, relation.target
    \State Normalize names of A and B using alias dictionary
    \State Get current position of B as $(x_b, y_b)$
    \State Compute new position $(x_a, y_a) \gets \text{ApplyOffset}(x_b, y_b, R)$
    \If{$(x_a, y_a)$ within bounds and not overlapping}
        \State Update A’s position in its assigned layer
    \EndIf
\EndFor
\end{algorithmic}
\end{algorithm}

The \texttt{ApplyOffset} function implements fixed spatial transformations:
\begin{itemize}
    \item \textbf{at the left of}: offset $(-3, 0)$
    \item \textbf{at the right of}: offset $(+3, 0)$
    \item \textbf{above}: offset $(0, -3)$
    \item \textbf{below}: offset $(0, +3)$
    \item \textbf{on top of}: overlapping placement (same coordinates but layer shift)
\end{itemize}


This symbolic-to-spatial grounding process supports interpretable scene composition and lays the foundation for future rule learning or agent-driven placement strategies.





\subsection{Knowledge Graph Construction and Narrative Linking}

To support symbolic reasoning, we construct optional scene-level knowledge graphs (KGs) derived from parsed narrative predicates. Each KG encodes the symbolic structure of a single story frame, while temporal relations across scenes are captured using a merged KG with `\texttt{precedes}` edges. These structures facilitate interpretable alignment between visual scenes and underlying narrative logic.

\subsubsection{Narrative Knowledge Graphs for Scene Composition}

\begin{figure*}[t]
  \centering
  \begin{subfigure}[t]{0.32\textwidth}
    \centering
    \includegraphics[width=\textwidth]{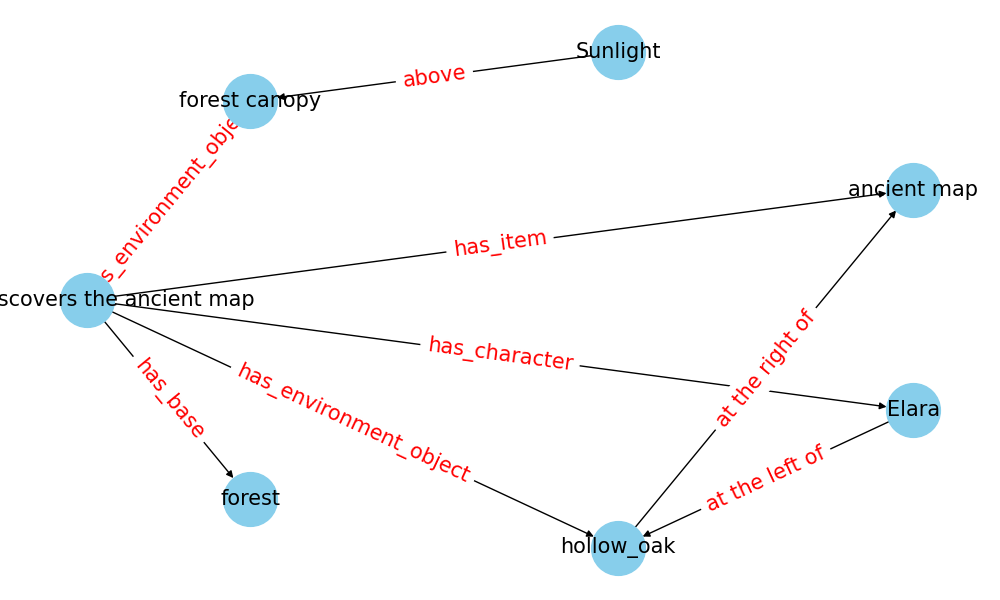}
    \caption{Elara discovers the ancient map}
    \label{fig:kg_frame1}
  \end{subfigure}
  \hfill
  \begin{subfigure}[t]{0.32\textwidth}
    \centering
    \includegraphics[width=\textwidth]{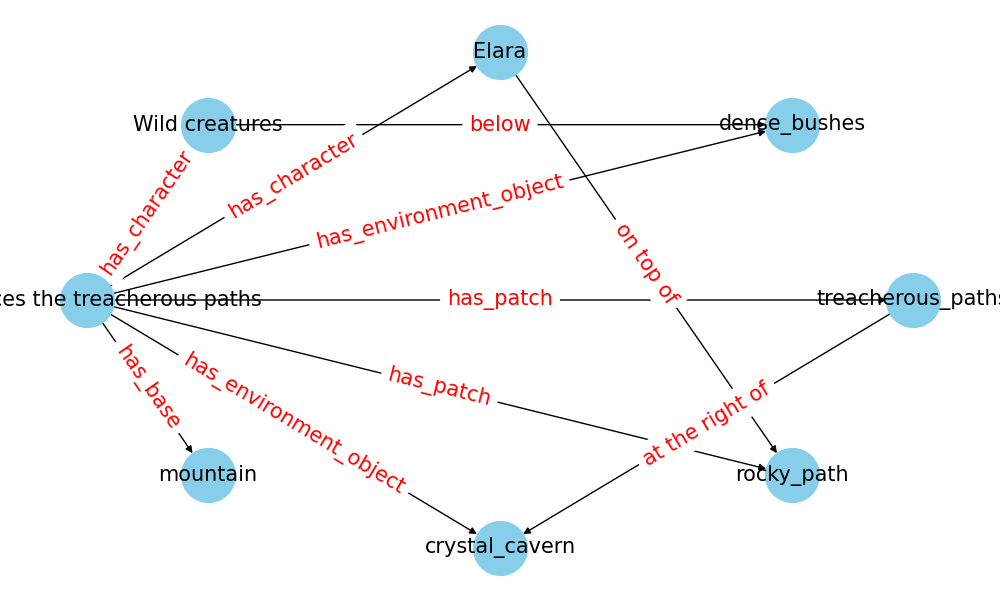}
    \caption{Elara faces the treacherous paths}
    \label{fig:kg_frame2}
  \end{subfigure}
  \hfill
  \begin{subfigure}[t]{0.32\textwidth}
    \centering
    \includegraphics[width=\textwidth]{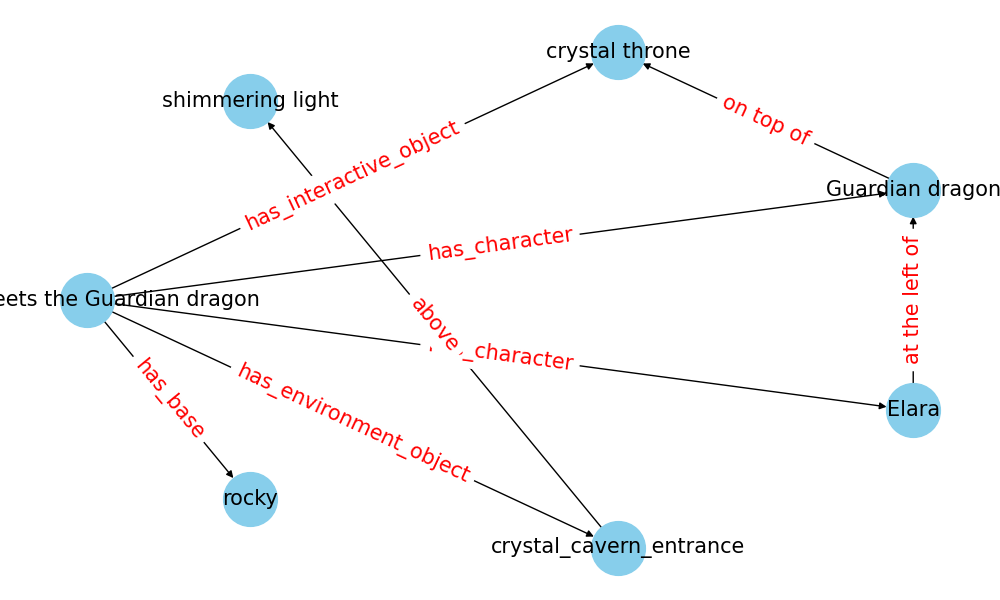}
    \caption{Elara meets the Guardian dragon}
    \label{fig:kg_frame3}
  \end{subfigure}
  \caption{Scene-level knowledge graphs capturing symbolic structure within three narrative frames.}
  \label{fig:story_kg_trio}
\end{figure*}

Each story scene is parsed into a set of symbolic predicates, typically in the form of \texttt{[Subject] [Relation] [Object]}. These predicates are transformed into triplets and rendered as a directed symbolic graph. Nodes represent objects and agents, while edges encode spatial or interactional relationships derived from language. Figure~\ref{fig:story_kg_trio} shows three examples of such graphs aligned with key scenes.

The knowledge graph for each frame includes:
\begin{itemize}
    \item \textbf{Entities:} Matched visual objects and characters.
    \item \textbf{Relations:} Symbolic spatial predicates (e.g., \texttt{above}, \texttt{contains}) derived from narrative text.
    \item \textbf{Semantic Roles:} Directional links such as agent-action-object when applicable.
\end{itemize}

These scene-level KGs enable localized symbolic reasoning and enhance traceability in narrative visualization.

\subsubsection{Cross-Frame Temporal Integration}

To preserve the overarching story flow, we link each scene-level KG through a unified structure. This \textit{merged knowledge graph} introduces \texttt{precedes} edges to connect scenes according to narrative chronology. These temporal links enable:
\begin{itemize}
    \item Global queries across multiple scenes.
    \item Timeline reconstruction.
    \item Coherence analysis across disconnected predicates.
\end{itemize}

When no scene boundary is detected in the narrative (i.e., no setting or goal shift), the system assumes the location is continuous. This continuity influences both terrain rendering and the KG linkage.

\subsubsection{Relation to Hierarchical Narrative Models}

Our construction is inspired by prior work on hierarchical symbolic representation for visual narratives~\cite{chen2025hierarchical}. In that model, events are represented at multiple levels, from panel to event to macro-event, and integrated using graph structures that encode semantic, temporal, and multimodal relations.

Although our current system focuses on symbolic graphs from text and image grounding, its scene-level KG resembles the \textit{event segment} layer in~\cite{chen2025hierarchical}. Both:
\begin{itemize}
    \item Represent discrete narrative units grounded in key scenes or moments.
    \item Encode semantic roles and inter-entity relations symbolically.
    \item Support integration into larger temporal and narrative structures.
\end{itemize}


While our current scene graphs are simpler, they lay the groundwork for symbolic extensions and integration with event-segment and macro-event abstractions. This future work could enable a unified narrative reasoning framework for both scene generation and story understanding.





\subsection{Visual Rendering and Output}

After semantic matching and spatial placement, each tile-based scene is rendered using matched 2D sprite assets. Scenes are organized into multi-layer matrices representing different object types (terrain, environment, interactive objects, items, characters), which are composited in semantic order to preserve depth and spatial logic. Object images are resized (e.g., $1.5\times$), centered within their tiles, and pasted layer by layer to construct the final image. Figure layers are rendered from background to foreground based on their affordance class. We also retain each layer's numerical matrix for downstream tasks such as symbolic reasoning or gameplay simulation. The rendering procedure is outlined in Algorithm~\ref{alg:render_scene}.

\vspace{0.5em}
\begin{algorithm}[H]
\caption{RenderSceneImage}
\label{alg:render_scene}
\begin{algorithmic}[1]
\Require scene\_layers, matched\_objects, scene\_summaries
\Ensure Saved visual rendering of each scene
\ForAll{scene in scene\_summaries}
    \State Initialize canvas with white background
    \State Draw base terrain using binary mask
    \ForAll{layer\_type in \{environment, interactive, item, character\}}
        \State Get object list and placement matrix
        \ForAll{(x, y), object in matrix positions}
            \State Lookup matched object image
            \State Resize and center image on tile
            \State Paste image onto canvas
        \EndFor
    \EndFor
    \State Save canvas as PNG to output folder
\EndFor
\end{algorithmic}
\end{algorithm}

\section{Evaluation}


We evaluated our narrative-to-scene generation system using 10 LLM-generated stories. Each story is segmented into three key time frames, resulting in a total of 30 scene visualizations and their associated symbolic representations. We assess the quality of the generated outputs from multiple perspectives: tile–object semantic alignment, affordance-layer placement correctness, spatial predicate satisfaction, and qualitative renderings. 

\subsection{Experimental Setup}

Each narrative (approximately 100 words) is decomposed into three time frames via prompting.  
For each frame, we extract three predicate triples, which are matched to GameTileNet assets using semantic embeddings and placed within procedurally generated terrains. Table~\ref{tab:dataset_summary} summarizes the evaluation dataset.

\begin{table}[h!]
\centering
\caption{Dataset summary.}
\label{tab:dataset_summary}
\begin{tabular}{cccc}
\toprule
\textbf{Stories} & \textbf{Scenes/Story} & \textbf{Entities/Scene} & \textbf{Total Scenes} \\
\midrule
10 & 3 & 4--6 & 30 \\
\bottomrule
\end{tabular}
\end{table}


\subsection{Tile Matching Accuracy}

We first examine whether the tiles selected by semantic matching correspond well to the narrative objects. Evaluation considers (1) whether the top-1 matched tile is semantically appropriate, (2) whether the assigned affordance matches the expected gameplay role, and (3) whether the system produces a diverse set of tiles across each story. Per-story results are shown in Table~\ref{tab:per-story-results}, and aggregate results across all 30 scenes are given in Table~\ref{tab:tile-aggregate}.


\begin{table}[h!]
\centering
\caption{Per-story evaluation results. CosSim: top-1 cosine similarity; Afford: affordance match rate; Div: diversity (1.0 = all unique); Sat: spatial predicate satisfaction rate. Each scene includes 3 predicates on average.}
\label{tab:per-story-results}
\begin{tabular}{lcccc}
\toprule
\textbf{Story} & \textbf{CosSim} & \textbf{Afford} & \textbf{Div} & \textbf{Sat (\%)} \\
\midrule
1  & 0.43 & 0.45 & 0.91 & 78 \\
2  & 0.40 & 0.33 & 0.87 & 67 \\
3  & 0.38 & 0.55 & 1.00 & 67 \\
4  & 0.41 & 0.36 & 1.00 & 67 \\
5  & 0.44 & 0.43 & 1.00 & 78 \\
6  & 0.43 & 0.45 & 0.82 & 89 \\
7  & 0.37 & 0.36 & 1.00 & 67 \\
8  & 0.41 & 0.27 & 0.82 & 78 \\
9  & 0.42 & 0.50 & 0.90 & 78 \\
10 & 0.44 & 0.54 & 0.85 & 56 \\
\midrule
Overall & 0.41 & 0.42 & 0.92 & 72 \\
\bottomrule
\end{tabular}
\end{table}

\begin{table}[h!]
\centering
\caption{Aggregate tile matching results (10 stories, 30 scenes).}
\label{tab:tile-aggregate}
\begin{tabular}{lcc}
\toprule
\textbf{Metric} & \textbf{Mean} & \textbf{Std. Dev.} \\
\midrule
Cosine similarity & 0.41 & 0.02 \\
Affordance match  & 0.42 & 0.09 \\
Diversity         & 0.92 & 0.07 \\
\bottomrule
\end{tabular}
\end{table}

\paragraph{\textbf{Analysis.}} 
The results in Table~\ref{tab:tile-aggregate} show that semantic alignment between narrative objects and candidate tiles is reliable across stories: cosine similarity values are consistently around 0.40--0.44 with low variance (Table~\ref{tab:per-story-results}). This suggests that narrative embeddings provide a stable signal for selecting visually coherent tiles. By contrast, affordance match rates fluctuate more strongly (0.27--0.55, mean 0.42), indicating that while visual semantics are captured, gameplay functions (e.g., terrain vs.\ item vs.\ obstacle) are more difficult to preserve. Tile diversity remains high across all stories (mean 0.92), showing that the system avoids reusing the same assets excessively and maintains scene variety. Error inspection revealed three recurring challenges: inconsistent naming conventions (e.g., “decrepit library” vs.\ “decrepit\_library”), coverage gaps where no suitable tile existed, and affordance misclassifications when a visually similar tile had the wrong role. Overall, these findings highlight the usefulness of semantic matching but also point to the need for stronger affordance-aware retrieval.

\subsection{Spatial Predicate Satisfaction}

We next examine whether spatial relations (e.g., “Tree to the left of House”) are satisfied in the rendered layout. A rule-based checker validates each predicate based on scene matrices, and we compute the percentage of predicates satisfied per scene. Results are shown in Table~\ref{tab:per-story-results}.


\paragraph{\textbf{Analysis.}} 
Predicate satisfaction averaged 72\% across all stories (Table~\ref{tab:per-story-results}), with the majority of scenes achieving two-thirds or more of their relational constraints. Stories 6 and 1 achieved the highest consistency (89\% and 78\%), while Story 10 lagged (56\%), typically due to conflicting placement constraints or lack of sufficient map space. These results suggest that the procedural layout is capable of enforcing basic spatial relations but can be brittle when multiple constraints interact. Improvements such as constraint-solving or affordance-informed placement could further enhance satisfaction rates.

\section{Discussion}


\subsection{Strengths and Generalization}

The evaluation highlights several promising aspects of our approach. 
First, semantic alignment between narrative descriptions and visual tiles is stable across all ten stories (Table~\ref{tab:tile-aggregate}), indicating that embedding-based retrieval provides a reliable foundation for mapping open-ended narrative text into game assets. 
Second, the system maintains high tile diversity (mean 0.92), suggesting that it can produce varied outputs without excessive repetition, an important property for replayability and player engagement. 
Third, spatial predicate satisfaction averaged 72\% (Table~\ref{tab:per-story-results}), demonstrating that even a lightweight rule-based layout generator can enforce a substantial fraction of narrative constraints. 
Together, these findings suggest that the pipeline generalizes across different story contexts, making it adaptable for varied game scenarios.

\subsection{Limitations}

Despite these strengths, several limitations remain. 
Affordance matching showed high variance (0.27–0.55), revealing a gap between visual similarity and gameplay semantics. This partly stems from limited coverage in the GameTileNet dataset: some objects (e.g., "lantern," "archway") lack representative tiles, forcing approximate matches. Semantic embeddings also capture descriptive but not functional similarity (e.g., terrain vs.\ collectible), leading to occasional misplacements. Symbolic layers were not fully leveraged for spatial reasoning, the layout engine checks constraints but does not resolve conflicts, causing brittle performance when multiple predicates interact. Finally, narrative-to-scene generation is inherently open-ended, complicating evaluation. Metrics such as diversity and cosine similarity depend not only on correct object interpretation but also on tile coverage and the ambiguity of natural language descriptions. The current three-frame segmentation follows a fixed narrative template; future versions will explore data-driven segmentation that adapts frame count to story complexity. Ablation studies on individual modules (terrain generation, semantic matching, spatial refinement) were not conducted; future work will quantify their contributions to overall scene coherence.

\subsection{Use Cases and Integration in Game Tools}

Despite these challenges, the framework has several promising use cases. 
For game developers, the system can serve as a prototyping tool, quickly transforming narrative prompts into playable scene sketches that can be refined by designers. 
For procedural content generation (PCG) research, it offers a testbed that integrates symbolic reasoning, semantic matching, and spatial layout, enabling controlled experiments on hybrid generation pipelines. 
Integration into existing game engines such as Unity or Godot could extend the system into interactive editors, where designers specify narrative beats and receive automatically generated candidate scenes. Beyond development, the pipeline may also support applications in game-based storytelling, educational games, or automated testing of narrative scenarios.

Overall, the results suggest that narrative-driven PCG is feasible, but requires a deeper integration of affordance-aware retrieval and constraint-solving methods to bridge the gap between narrative semantics and functional game design.

\section{Conclusion}

We presented a pipeline for generating game scenes from narrative text by aligning LLM-derived predicates with the GameTileNet dataset and rendering layered maps using procedural terrain generation. Our evaluation across ten stories demonstrated that semantic matching provides stable visual alignment with narrative objects, while affordance alignment and spatial relation enforcement remain challenging. These findings highlight both the promise of semantic embeddings for bridging text and assets and the need for deeper affordance-aware reasoning to ensure gameplay consistency. This work provides an early step toward narrative-driven procedural content generation. Future directions include integrating symbolic reasoning for more reliable spatial and temporal coordination, expanding the coverage of tile datasets to reduce gaps in representation, and supporting interactive or co-creative workflows where designers and players can iteratively refine generated scenes. We see these developments as important next steps toward practical tools that blend narrative expression with playable game environments.

\section*{Declaration on Generative AI}

Generative AI tools were used only to assist with language refinement and LaTeX formatting under the authors’ direction. All conceptual contributions, design, implementation, and analysis were produced by the authors.

\bibliography{bibfile}

\appendix

\vspace{-0.8em}
\section{Online Resources}
The Code and examples are released publicly via \href{https://github.com/RimiChen/2025_NarrativeScene}{https://github.com/RimiChen/2025\_NarrativeScene}.
\vspace{-0.8em}
\section{Appendix: Sample Stories}
\begin{tcolorbox}[colback=gray!5!white, colframe=gray!60!black, boxrule=0.5pt, arc=3pt]
\small
\textit{
  Amid the neon glow of New York’s restless night, journalist Jake stumbled upon a cryptic note tucked inside a trash can under a flickering streetlight. The message hinted at a hidden treasure buried deep within the city’s shadows. Pursued by ruthless gangsters, Jake raced through winding alleys bathed in moonlight, his every step echoing with danger. Clue after clue led him skyward—up spiral stairs and secret elevators—until he reached the crown of the Statue of Liberty. There, hidden beneath the cold iron floor, he uncovered the treasure. As dawn broke over the skyline, Jake realized he’d rewritten the city’s secrets.}
\end{tcolorbox}
\begin{tcolorbox}[colback=gray!5!white, colframe=gray!60!black, boxrule=0.5pt, arc=3pt]
\small
\textit{
In the neon-lit sprawl of Neo-Tokyo, young hacker Kenji uncovered a cache of encrypted files hidden in a derelict mainframe. Dust swirled in the glow of failing lights as lines of forgotten code blinked to life. Back in his cramped apartment, Kenji hunched over his terminal, fingers flying across the keyboard. The screen flooded with cascading symbols, then froze—decoding complete. What emerged was more than data; it was evidence of a vast corporate conspiracy. As city skyscrapers loomed outside his window, Kenji stared at the truth pulsing on his screen, knowing his next move could shake the world’s digital core.}
\end{tcolorbox}
\begin{tcolorbox}[colback=gray!5!white, colframe=gray!60!black, boxrule=0.5pt, arc=3pt]
\small
\textit{
In a crumbling library beneath a sagging ceiling, Iris unearthed a fragile message sealed in dust and time. The note whispered of a hidden oasis—a refuge in the desolate ruins of the world. With resolve burning in her chest, she ventured into the scorched wastelands, where mutant creatures prowled and the earth cracked beneath her feet. Days blurred into nights, but Iris pressed on. At the edge of collapse, she found it: a shimmering oasis blooming defiantly in the dead soil. As its waters sparkled with life, humanity’s hope rekindled. Iris hadn’t just survived—she had rediscovered a future.}
\end{tcolorbox}
\begin{tcolorbox}[colback=gray!5!white, colframe=gray!60!black, boxrule=0.5pt, arc=3pt]
\small
\textit{
Inside the public library, Alex stood near a dusty bookshelf, where an old book rested untouched. Opening it, he found it contained an encrypted note tucked between yellowed pages. That night, at home, Alex held the encrypted note over his desk, where moonlight illuminated the surface. The desk supported scattered papers, maps, and scribbled codes. After cracking the message, he followed its coordinates to an abandoned warehouse. There, Alex stood in the dim space, heart pounding. Streetlights outside cast shadows on the warehouse exterior. Suddenly, figures emerged—the secret society gathered around Alex, their eyes fixed on the note he still held.}
\end{tcolorbox}
\begin{tcolorbox}[colback=gray!5!white, colframe=gray!60!black, boxrule=0.5pt, arc=3pt]
\small
\textit{
In the rain-slicked alleys of Zephyr, streetwise Jax knelt by a loose cobblestone glowing faintly beneath the streetlight’s shimmer. Beneath it, he uncovered an emerald amulet pulsing with forgotten energy. Clutching it in his hand, a surge of ancient power surged through him, surrounding him in a halo of green fire. The city trembled. Above the skyline, a sinister sorcerer descended from the clouds. Jax stood his ground, the amulet blazing with newfound strength. Magic clashed in the sky, old and new. When the light faded, only Jax remained—victorious and forever changed by the artifact he had unearthed from the street.}
\end{tcolorbox}
\begin{tcolorbox}[colback=gray!5!white, colframe=gray!60!black, boxrule=0.5pt, arc=3pt]
\small
\textit{
Aboard the cursed ship The Sea Serpent, Captain Redbeard peered into the abyss, where the ocean floor cradled a glowing artifact of unknown origin. As the ship rocked atop the waves, he hauled it aboard, sensing the tide of fate shift. From the deep, monstrous shapes surged—leviathans with fangs like anchors. Redbeard stood firm, the artifact blazing in his grip as he battled the beasts. When the sea fell still, the relic rose, casting a vision across the sky: an ancient prophecy long forgotten. As its light danced across the waves, Redbeard knew the sea had chosen its next legend.}
\end{tcolorbox}
\begin{tcolorbox}[colback=gray!5!white, colframe=gray!60!black, boxrule=0.5pt, arc=3pt]
\small
\textit{
Amid a raging storm, Captain Jack stood firm on the deck, waves crashing around him. The sea surrounded his ship, The Stormcaller, as he gripped a cryptic compass unearthed from a sailor’s tale. Its needle trembled, pointing unerringly toward the fabled El Dorado. Navigating through the Bermuda Triangle, Jack’s vessel braved merciless waves while mystical sea creatures lunged from the depths. His crew fought with steel and fear. At last, the storm broke. Under radiant moonlight, the horizon cleared—revealing golden spires glistening in the distance. Standing on the drenched deck, Jack watched El Dorado rise from myth into reality.}
\end{tcolorbox}

\begin{tcolorbox}[colback=gray!5!white, colframe=gray!60!black, boxrule=0.5pt, arc=3pt]
\small
\textit{
In the blistering Sahara Desert, where the sun beats down on endless dunes, Dr. Samuel Cross knelt near an unearthed ancient amulet glinting in the sand. The Sahara Desert contained more than secrets—it held the path to legend. As he journeyed onward, a violent sandstorm engulfed Cross, tearing visibility to shreds. Through the storm, he read cryptic hieroglyphs etched in stone, while tomb raiders pursued him relentlessly. Deeper inside the pyramid, Cross stood in a hidden chamber—an elaborate pyramid trap that contained deadly mechanisms. Clutching the fabled treasure, Cross narrowly escaped, leaving behind danger but carrying history in his hands.}
\end{tcolorbox}
\begin{tcolorbox}[colback=gray!5!white, colframe=gray!60!black, boxrule=0.5pt, arc=3pt]
\small
\textit{
Deep within the abyss of the Forgotten City, intrepid archaeologist Dr. Alexander unearthed a mystical artifact concealed in an ancient tomb. Torchlight flickered across the ruins as he knelt beside the stone sarcophagus, uncovering secrets long buried. With the artifact in hand, he pressed forward, navigating a labyrinth of treacherous traps and walls that whispered forgotten chants. Guided by the artifact’s glow, he reached the heart of the tomb. There, the Guardian emerged, its form towering in silence. As chamber doors slammed shut, the artifact shimmered and activated a hidden mechanism. The Guardian bowed. Dr. Alexander had passed the test—and earned the city’s truth.}
\end{tcolorbox}
\begin{tcolorbox}[colback=gray!5!white, colframe=gray!60!black, boxrule=0.5pt, arc=3pt]
\small
\textit{
In the blazing Sahara, golden sunlight beamed on the golden amulet buried just beneath the sand. Amelia stood over the buried amulet, brushing away grains until it shimmered in full. She picked it up. As she followed the amulet's glow across the desert, sand dunes surrounded her. A venomous creature lurked behind one of the dunes, but she pressed on. Soon, she arrived at a hidden pyramid. Amelia stood before the ancient pyramid, awed by its size. With steady hands, she raised the amulet. It fit perfectly into the pyramid’s lock. Inside, the hidden pyramid held civilization’s long-lost secrets.}
\end{tcolorbox}

\end{document}